\documentclass[sigconf,nonacm]{acmart}

\AtBeginDocument{%
  }

\usepackage{pifont}

\usepackage{amsmath}
\usepackage{xurl}

\usepackage{booktabs} 
\usepackage{multirow}

\usepackage{graphicx} 

\usepackage{enumitem}

\usepackage{listings}
\usepackage{array}

\usepackage{algorithm}
\usepackage{algpseudocode}

\begin{document}
\title{MAP-Bench: Benchmarking Cross-Model Analytical Pipelines in Multi-Model Database Systems}

\author{Yuhan He}
\affiliation{%
  \institution{Northeastern University}
  \city{Shenyang}
  \country{China}}
\email{heyuhan.chn@foxmail.com}

\author{Tiezheng Nie}
\affiliation{%
  \institution{Northeastern University}
  \city{Shenyang}
  \country{China}}
\email{nietiezheng@mail.neu.edu.cn}

\author{Zihao Wen}
\affiliation{%
  \institution{Northeastern University}
  \city{Shenyang}
  \country{China}}
\email{wenzihao@mails.neu.edu.cn}

\author{Zhenyu Gou}
\affiliation{%
  \institution{Northeastern University}
  \city{Shenyang}
  \country{China}}
\email{gouzy@mails.neu.edu.cn}

\author{Derong Shen}
\affiliation{%
  \institution{Northeastern University}
  \city{Shenyang}
  \country{China}}
\email{shendr@mail.neu.edu.cn}

\author{Yue Kou}
\affiliation{%
  \institution{Northeastern University}
  \city{Shenyang}
  \country{China}}
\email{kouyue@cse.neu.edu.cn}

\author{Sheng Wang}
\affiliation{%
  \institution{Wuhan University}
  \city{Wuhan}
  \country{China}}
\email{swangcs@whu.edu.cn}

\author{Jiaheng Lu}
\affiliation{%
  \institution{University of Helsinki}
  \city{Helsinki}
  \country{Finland}}
\email{Jiaheng.lu@helsinki.fi}

\begin{abstract}
Multi-model database systems are evolving to natively support relational, document, graph, and increasingly vector data. This trend is driven by applications like Retrieval-Augmented Generation (RAG), which require cross-model analytical pipelines. However, existing multi-model benchmarks lack vector integration, rely on synthetic data struggling to preserve cross-model semantic consistency, and organize workloads by application domains rather than execution characteristics. To bridge this gap, we present \emph{MAP-Bench}, a benchmark for cross-model analytical pipelines over these data models. Derived from OpenAlex, an open scholarly dataset, MAP-Bench offers a semantically consistent dataset where entities are coherently represented across models and linked via global identifiers. Its workload suite organizes queries into four execution patterns, including hybrid-lookup, attribute-aggregation, vector-similarity, and graph-traversal to reflect realistic analytical workflows. Beyond latency, MAP-Bench also evaluates resource usage and query complexity. Results show that pattern-oriented workloads reveal distinct performance bottlenecks related to join strategies, parallelism, vector search placement, and graph traversal mechanisms. No single system dominates across all patterns, while polyglot orchestration requires an estimated 2–4$\times$ the query-writing effort of unified multi-model query languages.

\end{abstract}


\maketitle

\section{Introduction}

\begin{figure}[t]
    \centering
    \includegraphics[width=\linewidth]{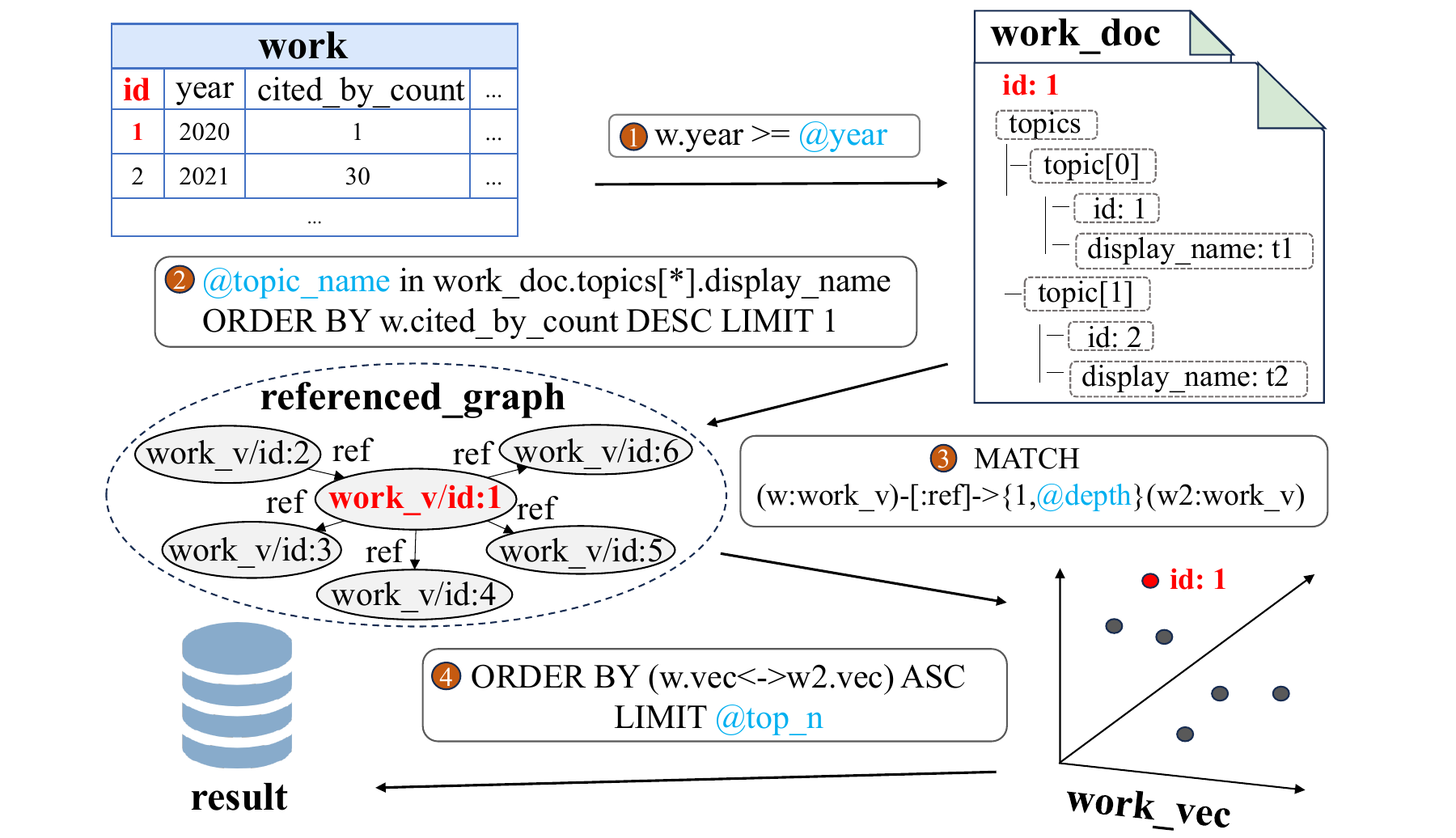}
    \caption{\textbf{Execution pipeline of a representative Graph-RAG retrieval query (Query V1) in MAP-Bench.}}
    \label{fig:pipeline}
\end{figure}

Multi-model data management is rapidly evolving, shifting from ``Polyglot Persistence''~\cite{DBLP:conf/hpec/GadepallyCDEHKM16}—integrating specialized engines for distinct data models—toward native \textbf{Multi-Model Databases (MMDBs)}~\cite{DBLP:journals/csur/LuH19}. Unlike fragmented polyglot architectures, MMDBs natively support relational, document, graph, and increasingly vector data within a single unified kernel. The rise of Generative AI~\cite{DBLP:conf/nips/VaswaniSPUJGKP17} accelerates this convergence, making vector embeddings essential for modern analytics~\cite{DBLP:journals/vldb/PanWL24}. Modern analytical workloads no longer operate in isolation; instead, they require joint execution across diverse data models to support advanced applications such as Retrieval-Augmented Generation (RAG)~\cite{DBLP:conf/nips/LewisPPPKGKLYR020}. For instance, a typical ``Graph-RAG''~\cite{DBLP:journals/corr/abs-2404-16130} query (Figure~\ref{fig:pipeline}) can involve identifying a seed paper filtered by year (Relational) and topic (Document), traversing its citation network (Graph), and finally recommending semantically similar papers via embeddings (Vector). Such complex, cross-model analytical pipelines are increasingly prevalent in modern search systems~\cite{DBLP:conf/kdd/HuangSSXZPPOY20} and enterprise analytics workloads~\cite{DBLP:journals/pvldb/ZhanSWPLWCLPZC19,DBLP:conf/er/HamadouGG19,DBLP:journals/pvldb/LeiLZZYXXWD24}, demanding architectures capable of efficiently unifying structured, semi-structured, topological, and semantic data.

This trend also changes how workloads are executed: they are often expressed as cross-model, multi-stage pipelines, where intermediate results flow across models.
Yet, the benchmarking ecosystem has not kept pace.
Existing benchmarks, such as UniBench~\cite{DBLP:conf/tpctc/ZhangLXC18} and M2Bench~\cite{DBLP:journals/pvldb/KimKEKKM22}, have limitations that restrict their suitability for modern cross-model analytical pipelines:

\begin{table*}[t]
\centering
\caption{Comparison of MAP-Bench with Representative Database Benchmarks.}
\label{tab:comparison}
\resizebox{\textwidth}{!}{%
\begin{tabular}{lccccc}
\toprule
\textbf{Benchmark} & \textbf{Target Models} & \textbf{Vector Integration} & \textbf{Semantic Consistency} & \textbf{Workload Focus / Organization} & \textbf{Key Metrics} \\ 
\midrule
\multicolumn{6}{l}{\textit{\textbf{Representative Single-Model Benchmarks}}} \\
\midrule
TPC-H / DS~\cite{tpch} & Rel & \ding{55} & N/A (Single Model) & Analytical Queries & Latency, Throughput \\
YCSB~\cite{DBLP:conf/cloud/CooperSTRS10} & KV/Doc & \ding{55} & N/A (Single Model) & Record-level CRUD & Latency, Throughput \\
LDBC SNB~\cite{DBLP:conf/sigmod/ErlingALCGPPB15} & Graph & \ding{55} & N/A (Single Model) & Interactive Graph Traversal & Latency, Throughput \\
VectorDBBench~\cite{vectordbbench_zilliztech} & Vec & \ding{51} (with scalar) & N/A (Single Model) & ANN \& Hybrid Search & Recall, Latency, Throughput \\
\midrule
\multicolumn{6}{l}{\textit{\textbf{Existing Multi-Model Benchmarks}}} \\
\midrule
UniBench~\cite{DBLP:conf/tpctc/ZhangLXC18} & Rel, Doc, Graph, KV & \ding{55} & Synthetic (Rule-based) & MM Queries \& Trans. (Business-level)  & Latency, Throughput \\
MDBench~\cite{DBLP:journals/jdm/YeSNSZ23} & Doc, Graph & \ding{55} & Synthetic (Rule-based) & Single/MM Ops (CRUD-categorized) & Throughput, Res., Availability \\
M2Bench~\cite{DBLP:journals/pvldb/KimKEKKM22} & Rel, Doc, Graph, Array & \ding{55} & Synthetic (Foreign-keys) & MM Analytics Tasks (Domain-specific) & Latency \\
\midrule
\addlinespace
\textbf{MAP-Bench (Ours)} & \textbf{Rel, Doc, Graph, Vec} & \textbf{\ding{51} (with other models)} & \textbf{Intrinsic (Real-world)} & \textbf{Multi-Model Pipelines (Pattern-oriented)} & \textbf{Latency, Res., Complexity} \\ 
\bottomrule
\end{tabular}%
}
\end{table*}

\begin{itemize}[leftmargin=*,nosep]
    \item \textbf{Absence of Vector Integration. } Existing multi-model benchmarks primarily focus on structured, semi-structured, and graph data. While standalone vector benchmarks~\cite{DBLP:journals/is/AumullerBF20,vectordbbench_zilliztech} evaluate the performance of vector similarity search in isolation, they do not capture its role within cross-model analytical workflows. Consequently, they cannot adequately represent emerging workloads like Graph-RAG that combine semantic retrieval with structural traversals.
    
     \item \textbf{Lack of Cross-Model Semantic Consistency.} Existing benchmarks lack a unified real-world multi-model dataset, and thus rely on synthetic mechanisms that struggle to preserve intrinsic cross-model semantic consistency, a key requirement for vector pipelines. UniBench derives other models unidirectionally from a pre-generated static graph, while M2Bench links independent datasets using synthetic foreign keys. These approaches do not fully capture the multi-model nature of real-world entities. For instance, a research paper naturally exists across multiple models—as relational metadata, document content, a citation graph, and vector embeddings—all of which are inherently correlated. When such correlations are not preserved, cross-model queries (e.g., retrieving papers that are both semantically similar to a given paper and connected in its citation graph) may yield inconsistent candidate sets, distorting intermediate result sizes and access patterns, and reducing workload realism.

    \item \textbf{Coarse-Grained Workload Organization.}
    Current multi-model benchmarks organize queries as flat lists (e.g., Q1–Qxx) or by high-level application domains (e.g., e-commerce) rather than by execution characteristics. This grouping does not explicitly distinguish differences in access patterns—such as selective lookups, scan-heavy aggregations, or recursive graph traversals—which exhibit fundamentally different execution structures. This makes it difficult to trace bottlenecks to the underlying query processing mechanisms, limiting the benchmark's utility for system tuning.
    
    \item \textbf{Single-Dimensional Evaluation Metrics.} Finally, existing evaluations overemphasize end-to-end latency, neglecting dimensions like resource consumption and query complexity. This narrow perspective obscures the true engineering cost, limiting a comprehensive assessment of both system efficiency and query development effort.
\end{itemize}

To bridge this gap, we present \textbf{MAP-Bench} (Multi-model Analytical Processing Benchmark), a benchmark for cross-model analytical pipelines over semantically consistent real-world data. It evaluates queries spanning relational (R), document (D), graph (G), and vector (V) models within a unified setting, capturing multi-stage analytical workflows across models. Built on OpenAlex~\cite{DBLP:journals/corr/abs-2205-01833}, an open scholarly dataset, MAP-Bench adopts a unified modeling strategy where each paper is consistently represented as structured metadata (R), nested content (D), citation topology (G), and semantic embeddings (V), linked by global identifiers to preserve cross-model semantic consistency important for realistic cross-model analytics. Table~\ref{tab:comparison} summarizes how MAP-Bench addresses these limitations compared to representative benchmarks.

While real-world snapshots ensure semantic consistency, ETL-based partitioning yields datasets at coarse, discrete scales. To enable fine-grained volume control, we offer an auxiliary simulation-based data generator driven by real OpenAlex statistics. By simulating key scholarly processes (e.g., citation, collaboration), the generator incrementally extends the dataset while consistently updating all related data models. As a result, this approach preserves the cross-model correlations in the original dataset, which are difficult to maintain with unidirectional or statically joined generation.

To systematically evaluate the performance of cross-model queries, we propose the \textbf{H-A-V-G} workload suite. The workloads are designed around four dominant query patterns: \textbf{H}ybrid-Lookup (Interactive Lookup), \textbf{A}ttribute-Aggre\-gation (Analytical Grouping and Ranking), \textbf{V}ector-Similarity (Semantic Candidate Retrieval), and \textbf{G}raph-Traversal (Structural Path Exploration).  Queries are implemented as multi-stage pipelines, where the output of one stage (e.g., vector search candidates) feeds into the next (e.g., graph traversal), simulating real-world exploratory analytics and RAG workflows. Furthermore, it provides a direct comparison of architectural approaches: for MMDBs, it measures the efficiency of internal data locality and unified query processing; for Polyglot systems, it quantifies the cost of data serialization and network transfers required to coordinate separate engines.

To move beyond single-dimensional metrics, we propose a holistic evaluation framework. In addition to latency, we evaluate \textbf{Performance Robustness} by comparing cold and hot executions, and measure \textbf{Resource Efficiency} (CPU time, peak memory, and average CPU cores) to capture runtime resource demand and parallelism utilization. Finally, we evaluate \textbf{Query Complexity} using Halstead metrics~\cite{halstead1977elements} to approximate the effort of writing cross-model queries. Beyond serving developers, query complexity also informs query language selection for emerging Text-to-Query automation (e.g., Text2SQL~\cite{DBLP:conf/emnlp/YuZYYWLMLYRZR18}). This exposes trade-offs between performance, resource usage, and query development effort.

In summary, this paper makes the following contributions:

\begin{itemize}[leftmargin=*,nosep]
    \item \textbf{Real-World Unified Dataset:} A native multi-model dataset derived from OpenAlex, with R/D/G/V models explicitly organized and linked via global identifiers to preserve authentic cross-model correlations (Section~\ref{sec:dataset}).

    \item \textbf{Scalable Data Extension:} An auxiliary, consistency-aware 
          data generator for fine-grained scale control, driven by real 
          OpenAlex statistics (Section~\ref{sec:generator}).

    \item \textbf{H-A-V-G Workload:} A workload suite organized by four 
      execution patterns (Hybrid-Lookup, Attribute-Aggregation, Vector-Similarity, and Graph-Traversal), 
      implemented as multi-stage pipelines to reflect diverse analytical patterns (Section~\ref{sec:workload}).

    \item \textbf{Holistic Evaluation Framework:} Multi-dimensional evaluation covering performance, resources, and complexity (Section~\ref{sec:framework}).
\end{itemize}

Our evaluation reveals distinct performance characteristics across query patterns, reflecting differences in query execution strategies, parallelism, and cross-model processing capabilities. No single system dominates all workloads, while unified multi-model languages tend to have lower query complexity.

\section{Related Work}
\label{sec:related_work}

We categorize existing work as single-model benchmarks, multi-model evaluation suites, and adjacent domains.

\textbf{Single-Model Database Benchmarks.} For relational databases, TPC-H~\cite{tpch} and TPC-DS~\cite{tpcds_spec} provide standardized analytical workloads. YCSB~\cite{DBLP:conf/cloud/CooperSTRS10} and its variants~\cite{DBLP:conf/cloud/PatilPRTXLGFR11} focus on record-level throughput and latency in cloud serving systems. For semi-structured data, XMark~\cite{DBLP:conf/vldb/SchmidtWKCMB02} and XBench~\cite{DBLP:conf/icde/YaoOK04} evaluate complex XQuery processing, while NOBENCH~\cite{DBLP:conf/webdb/ChasseurLP13} targets JSON attribute selection and aggregation, and DeepBench~\cite{DBLP:conf/dbtest-ws/Belloni0SR22} focuses on nested operations such as array unnesting and nested traversals. For graph data, LinkBench~\cite{DBLP:conf/sigmod/ArmstrongPBC13} evaluates storage performance via node-edge retrieval, LDBC-SNB~\cite{DBLP:conf/sigmod/ErlingALCGPPB15} covers transactional and analytical graph queries, and GooDBye~\cite{DBLP:conf/edbt/MatyjaszczykRW20} examines scale-out bottlenecks using industrial graph topologies. For vector search, ANN-Benchmarks~\cite{DBLP:journals/is/AumullerBF20} evaluates recall-latency trade-offs at the indexing-library level, whereas VectorDBBench~\cite{vectordbbench_zilliztech} measures ingestion and query performance at the vector database level, covering ANN and hybrid search with scalar predicates. However, both benchmarks evaluate vector search largely in isolation, without integrating vector retrieval into cross-model, multi-stage analytical pipelines.

\textbf{Multi-Model Database Benchmarks.}  UniBench~\cite{DBLP:conf/tpctc/ZhangLXC18} pioneered unified benchmarking by integrating relational, document (JSON/XML), graph, and key-value models in a social-commerce setting. Its unidirectional data generation derives other models from a static LDBC SNB~\cite{DBLP:conf/sigmod/ErlingALCGPPB15} social graph, potentially limiting the representation of complex cross-model dependencies. Its workloads are organized by business levels (e.g., individual, community), emphasizing cross-model joins and ACID transactions, while evaluation focuses mainly on execution latency. Another representative work, M2Bench~\cite{DBLP:journals/pvldb/KimKEKKM22}, broadens the scenarios by incorporating the Array model for scientific tasks (e.g., linear algebra). It links independent datasets via synthetic identifiers but scales each model in isolation. This hinders the preservation of consistent joint distributions, compromising statistical realism. Similar to UniBench, it organizes tasks by application domains (e.g., healthcare, disaster) rather than execution characteristics. MDBench~\cite{DBLP:journals/jdm/YeSNSZ23} extends UniBench to evaluate MMDB versus polyglot architectures, introducing metrics for resource consumption and system availability. However, it primarily targets Document and Graph models and inherits UniBench's rule-based generation, which limits intrinsic semantic consistency.

\textbf{Adjacent Benchmarking Domains.} Other benchmarks explore hybrid scenarios: TPC-DI~\cite{DBLP:journals/pvldb/PoessRC14} and BigBench~\cite{DBLP:conf/sigmod/GhazalRHRPCJ13} target heterogeneous data integration and mixed-format big data analytics, respectively. Galli et al.~\cite{DBLP:journals/is/BimonteGMR22} assess the impacts of logical modeling on multi-model warehouses, while FONT~\cite{DBLP:conf/icde/ShiWZW24} provides a flexible visual evaluation platform for polystore systems. 
MMDBench~\cite{DBLP:conf/bench/MaoHLWRWS23} focuses on multimedia retrieval via property graphs. Finally, RAG benchmarks like RGB~\cite{DBLP:conf/aaai/0011LH024} and CRAG~\cite{DBLP:conf/nips/YangSXSBCCGJJKM24} assess LLM generation quality, distinct from our focus on database execution efficiency.

Unlike prior benchmarks (Table~\ref{tab:comparison}), MAP-Bench integrates vector retrieval into pattern-oriented cross-model analytical pipelines over semantically consistent real-world data, enabling multi-dimensional evaluation.

\section{MAP-Bench Overview}

\label{sec:overview}
\begin{figure*}[t]
    \centering
    \includegraphics[width=0.75\textwidth]{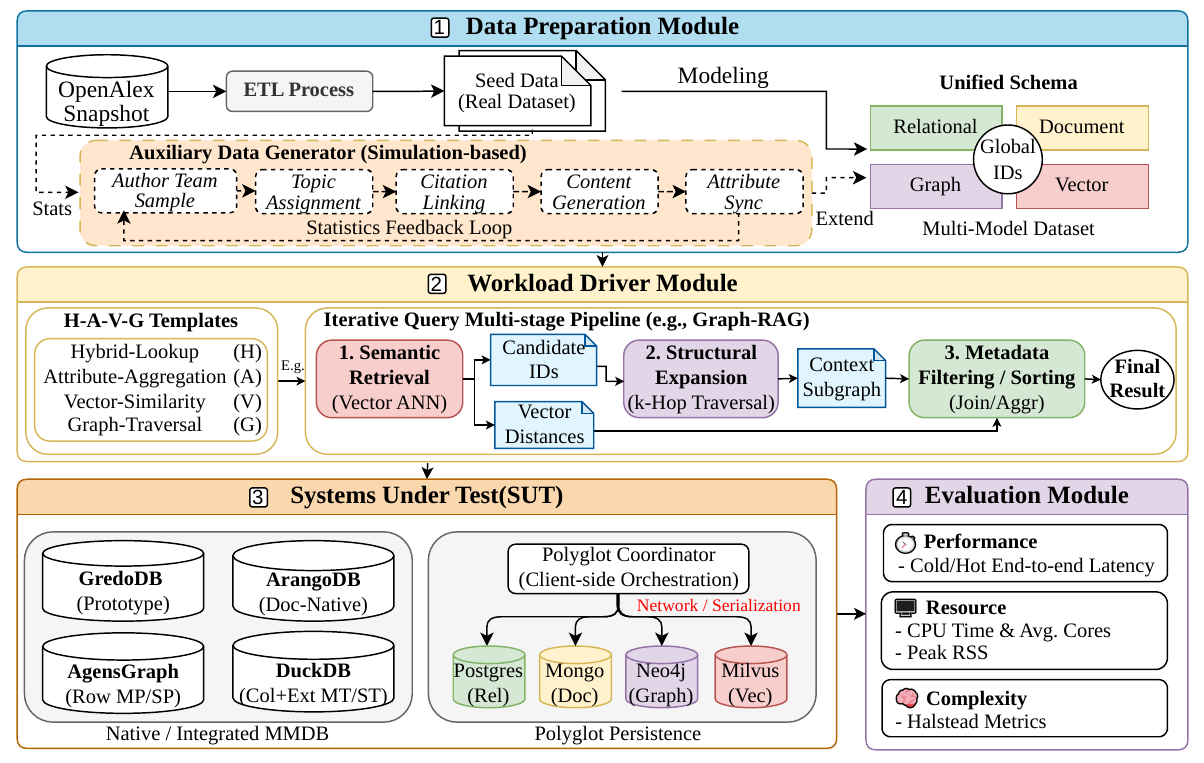}
    \caption{\textbf{MAP-Bench Architecture.} Modules: (1) Data Prep, (2) Workload Driver, (3) SUT Adapters, (4) Evaluation.}
    \label{fig:architecture}
\end{figure*}


Figure~\ref{fig:architecture} illustrates the MAP-Bench architecture, comprising four modules organized around the execution lifecycle of cross-model analytical queries: data preparation, workload orchestration, system adaptation, and evaluation.

The \textbf{Data Preparation Module} transforms a real-world OpenAlex snapshot into a unified multi-model dataset with global identifiers across models. This real-world dataset derived via ETL serves as the primary evaluation data, while an auxiliary simulation-based generator supports incremental data extension while preserving cross-model semantic consistency. The \textbf{Workload Driver Module} consumes this data to define workloads as logical multi-stage pipelines following the H-A-V-G templates. These pipelines specify the data dependencies between stages at a logical level, independent of the underlying system architecture.

The logical pipelines are implemented on the \textbf{Systems Under Test (SUTs)} via adapters handling data loading and physical query implementation. For MMDBs, a pipeline is expressed as a single query using constructs such as common table expressions (CTEs), enabling joint optimization and execution within one system. In contrast, polyglot architectures decompose the same pipeline into multiple queries coordinated by a client-side controller, incurring cross-engine data exchange. For fair comparison, workloads are implemented in each system's native dialect, following official best practices and approximately equivalent indexing strategies. All scripts are publicly available and serve as the basis for complexity analysis. Finally, the \textbf{Evaluation Module} records execution traces and system-level statistics, including latency, resource usage, and query complexity, for systematic comparison across architectures.

\section{The Unified Multi-Model Dataset}
\label{sec:dataset}
\begin{figure}[t]
    \centering
    \includegraphics[width=0.8\linewidth]{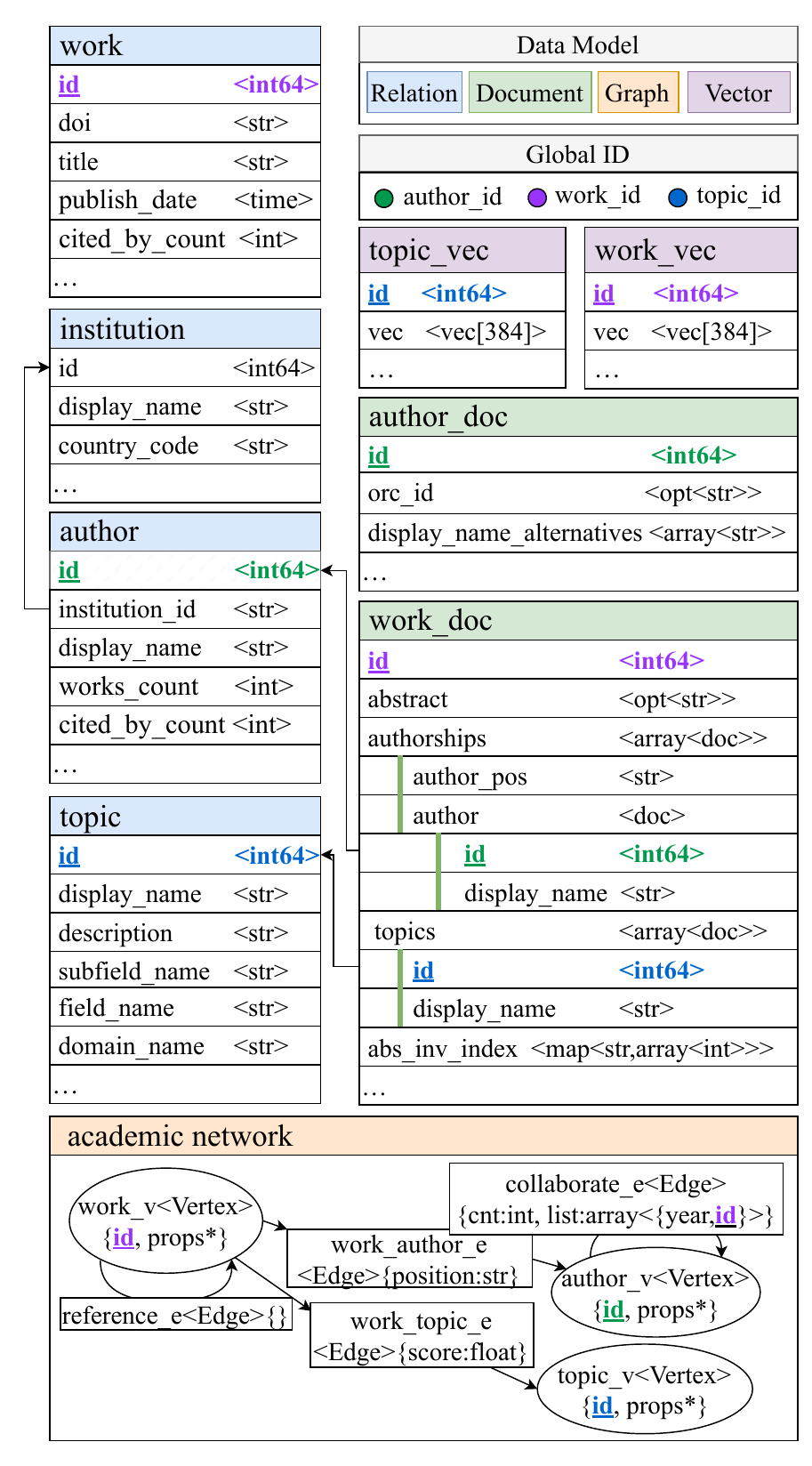}
    \caption{\textbf{MAP-Bench Schema.} Entities are projected across four models linked by Global IDs. Only representative attributes are shown. 
    In the Graph model, \texttt{props*} denotes vertex properties materialized for traversal filtering}
    \label{fig:schema}
\end{figure}

Emerging analytical pipelines often involve joint execution of multiple data models, 
where intermediate results from one model feed into subsequent stages in others. 
To support such workflows, a benchmark dataset should satisfy three design principles: 
(1) \textbf{semantic consistency across models}, ensuring real-world entities are represented coherently across different data representations; 
(2) \textbf{native support for multi-stage pipelines}, allowing intermediate results to flow coherently through explicit, semantically grounded cross-model correlations; and 
(3) \textbf{scalability}, supporting dataset growth while preserving these correlations. 
MAP-Bench achieves these goals through an entity-centric design based on OpenAlex~\cite{DBLP:journals/corr/abs-2205-01833}.

\subsection{Unified Schema Design}
\label{sec:schema_design}

Guided by the above design objectives, MAP-Bench adopts an entity-centric modeling strategy described as \textit{``One Entity, Multi-View''}. 
Real-world academic entities—primarily works, authors, and topics—are projected into four distinct data models anchored by ``Global Identifiers (GIDs)''. As illustrated in Figure~\ref{fig:schema}, identifiers such as \texttt{work\_id} and \texttt{author\_id} act as the primary join keys across the dataset, ensuring coherent cross-model representation and stable linkage for multi-stage pipelines (e.g., from vector search to graph traversal). The schema retains the structural features of each data model that are necessary for core operations.

The \textbf{Relational Model (R)} captures structured metadata defining core scholarly entities, characterized by compact schemas with frequently updated attribute values.
Tables such as \texttt{work} and \texttt{author} store fundamental descriptors (e.g., \texttt{title}) alongside statistical metrics (e.g., \texttt{cited\_by\_count}) that update with the scholarly record.
These entities are linked via explicit foreign-key relationships (e.g., \texttt{institution\_id}), following conventional normalized relational design.

The \textbf{Document Model (D)} targets cold, semi-structured content under a schema-on-read paradigm. Collections like \texttt{work\_doc} store variable-length text (e.g., \texttt{abstract}) and sparse attributes (e.g., \texttt{orc\_id}), avoiding the rigid constraints and null-storage overhead of relational tables. Crucially, this model preserves data locality for nested hierarchies, from simple arrays like \texttt{display\_name\_alternatives} to deep structures such as \texttt{author\-ships}. It also preserves the native \texttt{abs\_inv\_index} from OpenAlex (mapping terms to positions), which tests the system's efficiency in handling real-world schema-less patterns.

The \textbf{Graph Model (G)} represents complex scholarly relationships as a Property Graph~\cite{DBLP:conf/amw/Angles18}, supporting homogeneous and heterogeneous topologies. Vertices (e.g., \texttt{work\_v}, \texttt{author\_v}, \texttt{topic\_v}) retain filtering properties (e.g., \texttt{display\_name} from R-Model) and are linked by typed edges. Specifically, the schema includes homogeneous relations like the undirected \texttt{collaborate\_e} (for community analysis) and the directed \texttt{reference\_e} (citations), alongside heterogeneous relations like \texttt{work\_topic\_e} which support multi-hop traversals across different entity types (i.e., meta-paths~\cite{DBLP:journals/pvldb/SunHYYW11}). In addition, \texttt{collaborate\_e} is defined with nested properties, including a collaboration count and a list of co-authored works (annotated with \texttt{year} and \texttt{id}). This structure enables the evaluation of systems’ handling of nested properties on graph edges.

Finally, the \textbf{Vector Model (V)} stores high-dimensional semantic embeddings derived from paper abstracts and topic descriptions (e.g., \texttt{work\_vec}). In production systems, embeddings are typically generated by offline machine learning pipelines and periodically refreshed, independent of transactional data. Large-scale architectures often deploy embedding storage as independent services, decoupled from relational databases. MAP-Bench reflects this prevalent pattern by modeling vectors as distinct entities linked via global identifiers, ensuring the benchmark accurately represents real-world data organization and access patterns.

\subsection{Real-World Dataset}

MAP-Bench is entirely constructed from OpenAlex~\cite{DBLP:journals/corr/abs-2205-01833}, a large-scale, open, CC0-licensed catalog of the global scholarly ecosystem, widely adopted in bibliometric/scientometric studies~\cite{DBLP:journals/corr/abs-2404-17663,DBLP:journals/efi/KonchH26,DBLP:journals/corr/abs-2601-21908}.
Following the modeling strategy in Section~\ref{sec:schema_design},
we extract four core entities (work, author, topic, and institution) from OpenAlex to form a coherent dataset and project them into globally linked R/D/G/V models. We provide three preconfigured dataset configurations (Table~\ref{tab:dataset_stats}) curated using consistent temporal and domain criteria, with larger scales supported by our auxiliary data generator. For vectors, we generate 384-dimensional embeddings using all-MiniLM-L6-v2~\cite{DBLP:conf/emnlp/ReimersG19} on titles/abstracts and topic descriptions.

\begin{table}[h]
    \centering
    \caption{\textbf{MAP-Bench Dataset Statistics.} 
    \textmd{Unless noted, counts in Millions (M). }}
    \label{tab:dataset_stats}
    \vspace{-2mm}
    \resizebox{0.48\textwidth}{!}{%
    \begin{tabular}{lrrrrrr}
        \toprule
        \textbf{Dataset} & \textbf{Rel} & \textbf{Doc} & \textbf{Vec} & \textbf{Nodes} & \textbf{Edges} & \textbf{Size (GB)} \\
        \midrule
        \textbf{\textsc{MAP-S}}  & 1.25 & 1.21 & 0.27 & 1.22 & 12.32 & 5.11 \\
        \textbf{\textsc{MAP-M}} & 5.15 & 5.07 & 1.65 & 5.07 & 35.47 & 25.91 \\
        \textbf{\textsc{MAP-L}}  & 13.54 & 13.28 & 5.82 & 13.28 & 89.10 & \textbf{95.24} \\
        \bottomrule
    \end{tabular}%
    }
    \vspace{-2mm}
\end{table}

While grounded in academic data, MAP-Bench targets common system-level bottlenecks. Its power-law graph topology introduces realistic data skew, while nested documents impose non-trivial parsing and storage overheads. By composing fundamental operations such as index lookups, graph traversals, and aggregations across models, MAP-Bench is built around workload patterns also seen in modern multi-model analytical applications (e.g., RAG and recommendation), providing a basis for studying similar patterns beyond the specific dataset.

\subsection{Scalable Data Extension}
\label{sec:generator}

\begin{figure}[t]
    \centering
    \includegraphics[width=\linewidth]{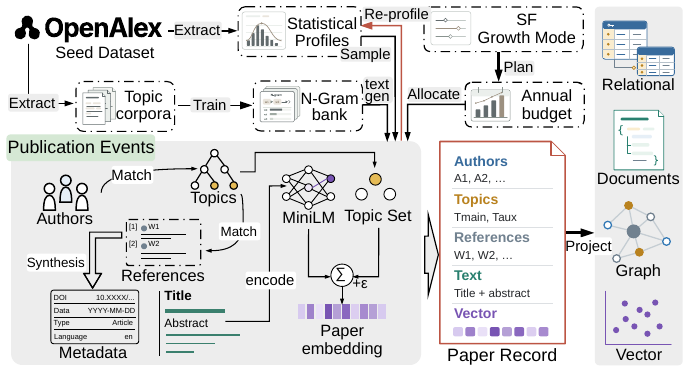}
    \caption{\textbf{Data Generator.} }
    \label{fig:gen}
\end{figure}

MAP-Bench uses real-world datasets derived from OpenAlex for evaluation. However, these ETL-derived datasets typically have fixed sizes determined by the extraction pipeline, limiting arbitrary scaling. To address this limitation, MAP-Bench provides an auxiliary data generator that preserves the attribute distributions and structural characteristics of the original data while maintaining semantic consistency across data models, as illustrated in Figure~\ref{fig:gen}.

The generator first characterizes the distributional properties of the seed dataset and constructs a topic corpus to train an n-gram model for text generation. Text generated by the n-gram model preserves a degree of semantic information while avoiding the substantial time overhead associated with LLM-based generation. Meanwhile, as statistical characteristics such as research topic popularity may continuously change with scholarly activities, the generator introduces a periodic statistical feedback mechanism that re-estimates relevant characteristics based on the accumulated generated data to guide subsequent event simulations. 

We adopt an event-driven generation approach, defining each paper publication as a basic event and incrementally extending the dataset through multiple rounds of event simulation. Within each event, the generator defines author activity and selects an author team based on historical collaboration relationships. The research topics are jointly determined by the research areas of the team members, while references are sampled based on domain relevance, citation statistics, and temporal constraints, thereby establishing the entities involved in the event and their topological relationships. Entity metadata are sampled from the statistical distributions of the seed dataset, long-form text is generated using an n-gram model, and vector representations are constructed using MiniLM based on the generated text and topic information.

Since the same entity may have different representations across data models, independently generating data for each model may compromise cross-model semantic consistency. Therefore, we organize the data produced by each event into a unified ``paper record'' and then map it to the relational, document, graph, and vector models according to the unified MAP-Bench schema, thereby preserving the inherent relationships across models and maintaining cross-model data consistency.

\section{The H-A-V-G Workload Suite}
\label{sec:workload}
\begin{figure*}[t]
    \centering
    \definecolor{keyblue}{rgb}{0.0, 0.3, 0.7}
    \definecolor{comgreen}{rgb}{0.0, 0.5, 0.0}
    \definecolor{strpurple}{rgb}{0.5, 0.0, 0.5}
    \definecolor{backgray}{rgb}{0.98, 0.98, 0.98}
    \definecolor{header}{rgb}{0.0, 0.0, 0.9}
    \definecolor{paramorange}{rgb}{1.0, 0.4, 0.0}

    \lstset{
        language=SQL,
        backgroundcolor=\color{backgray},
        basicstyle=\fontsize{5}{5}\selectfont\ttfamily, 
        keywordstyle=\color{keyblue}\bfseries,
        commentstyle=\color{comgreen}\itshape,
        stringstyle=\color{strpurple},
        morekeywords={MATCH, RETURN, WITH, UNNEST, LATERAL, ORDER, BY, LIMIT, CROSS, JOIN, WHERE, SELECT, FROM, GROUP, ON, AS, IN, OVER, PARTITION, DISTINCT, AND, OR, NOT, CASE, END},
        breaklines=true,
        breakindent=2pt,
        frame=single, 
        framerule=0pt, 
        framesep=5pt, 
        rulecolor=\color{black!10},
        aboveskip=0pt,
        belowskip=0pt,
        showstringspaces=false,
        tabsize=2,
        captionpos=b,
        escapechar=|,
        alsoletter=@,
        keywordstyle=[2]\color{paramorange},
        morekeywords=[2]{@year,@top_n,@topic,@depth,@cnt,@author_id,@old_name},
        moredelim=*[l][\color{header}\bfseries]{---},
    }

    \newsavebox{\codeH}
    \newsavebox{\codeA}
    \newsavebox{\codeV}
    \newsavebox{\codeG}

    \begin{lrbox}{\codeH}
    \begin{minipage}[t]{0.235\textwidth}
\begin{lstlisting}
--- STAGE 1: JSON Filter (Doc)
-- Find authors by old name
WITH Authors AS ( 
  SELECT id FROM author_doc 
  WHERE json_contains(
    display_name_alternatives,
    @old_name) 
)
--- STAGE 2: Context (Graph)
-- 1-hop traversal
SELECT a.id, a.display_name,
       g.titles, g.cnt
FROM (
  MATCH (au:author_v)
    <-[:work_author_e]-(w)
  WHERE au.id IN Authors
  WITH au.id AS id, 
    collect(w.title) AS titles,
    count(w.title) AS cnt
  RETURN id, titles, cnt
) g 
JOIN author a ON a.id = g.id;
\end{lstlisting}
    \end{minipage}
    \end{lrbox}

    \begin{lrbox}{\codeA}
    \begin{minipage}[t]{0.235\textwidth}
\begin{lstlisting}
--- STAGE 1: Unwind & Agg (Rel+Doc)
-- Count per institution
WITH TopInst AS (
  SELECT ship.inst.id
  FROM work w
  JOIN work_doc wd ON ...
  CROSS JOIN UNNEST(
    wd.authorships) AS ship
  WHERE ship.inst.id IS NOT NULL 
    AND w.year >= @year
  GROUP BY ship.inst.id
  ORDER BY count(1) DESC
  LIMIT @top_n
),
--- STAGE 2: Pattern Matching (Graph)
-- Map inst to topics
InstFields AS (
  SELECT g.i_id, g.t_id, 
         count(1) AS freq
  FROM (
    MATCH (au)<-[:work_author_e]
          -(w)-[:work_topic_e]
          ->(t)
    WHERE au.inst_id IN TopInst 
      AND w.year >= @year
    RETURN au.inst_id AS i_id,
           t.id AS t_id
  ) g
  GROUP BY g.i_id, g.t_id
)
--- STAGE 3: Window Rank (Rel)
SELECT i.display_name, t.display_name,
       k.freq, k.rank
FROM (
  SELECT inf.i_id, inf.t_id, inf.freq,
    row_number() OVER (
      PARTITION BY inf.i_id 
      ORDER BY inf.freq DESC
    ) AS rank
  FROM InstFields inf
  WHERE rank <= 3
) k
JOIN institution i ON ...
JOIN topic t ON ...;
\end{lstlisting}
    \end{minipage}
    \end{lrbox} 

    \begin{lrbox}{\codeV}
    \begin{minipage}[t]{0.235\textwidth}
\begin{lstlisting}
--- STAGE 1: Meta Filter (Rel+Doc)
-- Seed via scalar/JSON
WITH TopCitedPaper AS (
  SELECT w.id AS wid
  FROM work w
  JOIN work_doc wd ON ...
  WHERE w.year >= @year
    AND @topic IN 
      wd.topics[*].display_name
  ORDER BY w.cited_by_count DESC
  LIMIT 1
),
--- STAGE 2: Context (Graph)
-- Multi-hop citation net
CiteNet AS (
  MATCH (p1:work_v)
    -[:reference_e*1..@depth]
    ->(p2)
  WHERE p1.id = TopCitedPaper.wid
  RETURN DISTINCT p2.id AS cid
),
--- STAGE 3: Semantic Rank (Vector)
-- Rank by vector dist
SimScore AS (
  SELECT id AS cid,
    dist(vec,
      (SELECT vec FROM work_vec 
       WHERE id = TopCitedPaper.wid)
    ) AS score
  FROM work_vec 
  WHERE id IN CiteNet
  ORDER BY score ASC
  LIMIT @top_n
) 
--- STAGE 4: Final Projection
SELECT w.title, ss.score
FROM SimScore ss
JOIN work w ON ...
ORDER BY ss.score ASC;
\end{lstlisting}
    \end{minipage}
    \end{lrbox}

    \begin{lrbox}{\codeG}
    \begin{minipage}[t]{0.235\textwidth}
\begin{lstlisting}
--- STAGE 1: Topic lookup (Rel)
-- Get topic vector
WITH TopicInfo AS (
  SELECT t.id, tv.vec
  FROM topic t 
  JOIN topic_vec tv ON ...
  WHERE t.display_name = @topic
  LIMIT 1
),
--- STAGE 2: Deep Path & Prune (Graph)
-- 2-4 hops, exclude direct
Potential AS (
  MATCH (me)-[:collaborate_e*2..4]
        -(cand)
  WHERE me.id = @author_id 
    AND cand.cited_by_count >= @cnt 
    AND NOT (me)-[:collaborate_e]
            ->(cand)
  RETURN DISTINCT cand.id 
),
--- STAGE 3: Hybrid Filter (Doc+Graph)
-- Candidates in topic
CandWork AS (
  SELECT t.aid, wd.id
  FROM (
    MATCH (au)<-[:work_author_e]-(w)
    WHERE au.id IN Potential
    RETURN au.id AS aid, w.id
  ) t 
  JOIN work_doc wd ON ...
  WHERE TopicInfo.id IN 
        wd.topics[*].id
),
--- STAGE 4: Recommendation (Rel+Vector)
SELECT cw.aid,
  sum(1.0 / (1.0 + 
    dist(wv.vec, TopicInfo.vec))) AS score
FROM CandWork cw 
JOIN work_vec wv ON ...
GROUP BY cw.aid
ORDER BY score DESC;
\end{lstlisting}
    \end{minipage}
    \end{lrbox}

    \resizebox{\textwidth}{!}{%
        \begin{tabular}{
        >{\centering\arraybackslash}p{0.24\textwidth}
        >{\centering\arraybackslash}p{0.24\textwidth}
        >{\centering\arraybackslash}p{0.24\textwidth}
        >{\centering\arraybackslash}p{0.24\textwidth}
        }
        \setlength{\tabcolsep}{1pt}
            {\small\bfseries (a) H1: Hybrid-Lookup} &
            {\small\bfseries (b) A1: Attribute-Aggregation} &
            {\small\bfseries (c) V1: Vector-Similarity} &
            {\small\bfseries (d) G1: Graph-Traversal} \\[-0.45cm]
            \vtop{\null\centering\usebox{\codeH}} &
            \vtop{\null\centering\usebox{\codeA}} &
            \vtop{\null\centering\usebox{\codeV}} &
            \vtop{\null\centering\usebox{\codeG}}
        \end{tabular}%
    }
    \caption{\textbf{SQL-Like Pseudocode Illustrating the Logic of Representative H-A-V-G Queries in MAP-Bench.}
    Modeled as multi-stage pipelines, these queries target:
    (a) \textsc{find} papers by author name variants,
    (b) \textsc{analyze} topic distribution of top institutions,
    (c) \textsc{recommend} semantically similar papers within a citation graph, and
    (d) \textsc{explore} potential collaborators based on co-authorship and topic relevance.}
    \label{fig:sql_patterns}
\end{figure*}

In this section, we introduce the H-A-V-G workload suite, a collection of cross-model analytical queries structured around four dominant execution patterns.
\subsection{Workload Design}

The workload is designed to capture diverse system behaviors using representative cross-model queries. Our design methodology is guided by three principles:
\textbf{(1) Execution Pattern Classification.} Rather than grouping workloads by application domain, we organize them by dominant execution patterns (H-A-V-G). This decouples workloads from business semantics and highlights distinct performance characteristics, such as index seek efficiency, analytical scan throughput, structural traversal latency, and high-dimensional similarity search.

\textbf{(2) Holistic Operator Coverage.} 
Queries exercise representative operators across all four data models: relational joins and window-based aggregation (R), nested JSON extraction and containment predicates (D), multi-hop traversal and shortest-path search (G), and vector similarity computation with ANN search (V). This ensures that both single-model and cross-model behaviors are adequately evaluated (operator mapping detailed in the public artifact repository).

\textbf{(3) Multi-Stage Pipelining.} To reflect realistic analytical workflows and stress cross-model data movement in unified and polyglot architectures, queries are implemented as multi-stage pipelines, ranging from shallow chains to deeper pipelines with branching stage dependencies.

\subsection{Workload Patterns}
\label{sec:pattern}
To systematically evaluate cross-model analytical capabilities, MAP-Bench organizes workloads into four patterns: Hybrid-Lookup (H), Attribute-Aggregation (A), Vector-Similarity (V), and Graph-Traversal (G), whose representative queries are shown in Figure~\ref{fig:sql_patterns}. These patterns are inspired by the LDBC SNB workload classes (IS/IC/BI)~\cite{DBLP:conf/sigmod/ErlingALCGPPB15}, as well as real-world academic search and recommendation scenarios~\cite{googlescholar,dblp,DBLP:conf/wsdm/0001A26,DBLP:journals/corr/abs-2512-06879,DBLP:journals/kais/ZhangPYZSRCW23}. While applying these concepts to multi-model environments that include cross-model and vector operations, each pattern is designed to emphasize a characteristic query objective and its associated execution behavior. This approach provides a practical way to ensure workload diversity while preserving realistic multi-model composition.

The \textbf{Hybrid-Lookup (H) pattern} characterizes interactive, high-selectivity workloads, representative of real-time profile page rendering or entity detail retrieval. Queries typically start from a specific identifier (e.g., author name) to retrieve a small subgraph or nested document fields. This pattern is designed to stress random I/O latency and index efficiency, as the engine must efficiently coordinate access across heterogeneous data structures—such as combining inverted index lookups with graph adjacency traversals. For example, Query H1 first finds authors via a nested JSON search and then navigates their publication graph to fetch titles, challenging the system's ability to minimize random I/O overhead during cross-model data retrieval.

The \textbf{Attribute-Aggregation (A) pattern} focuses on analytical reporting, such as trend analysis or business intelligence. These queries scan large volumes of multi-model data to flatten, group, and sort records. Performance in this pattern is often constrained by memory bandwidth and the CPU cost of schema-on-read parsing, especially when unnesting complex JSON structures. For example, Query A1 analyzes top research institutions by scanning millions of documents, unnesting authorship arrays, and ranking research fields using graph pattern matching and window functions. This query exposes the cost of aggregations and cross-model joins over heterogeneous data models.

The \textbf{Vector-Similarity (V) pattern} integrates semantic retrieval into multi-model analytical workflows. Queries typically employ vector similarity search to identify relevant candidates, combined with relational filters, graph traversals, and document lookups. Performance reflects the system's ability to coordinate vector computations with other data models rather than standalone search efficiency. For example, Query V1 selects a seed paper via relational and document filters, explores its citation network, and recommends candidates by vector similarity. This workflow evaluates the engine's efficiency in incorporating vector retrieval within a multi-stage, cross-model pipeline.

The \textbf{Graph-Traversal (G) pattern} focuses on complex structural exploration, matching topological patterns such as shortest paths or multi-hop subgraphs, and integrating cross-model data. Queries perform recursive navigation across multiple hops. This pattern highlights the effects of random memory access and tests the efficiency of path enumeration. For example, Query G1 explores potential collaborators within a 2–4 hop range in the co-authorship network, excluding direct connections, and ranks candidates using vector distances. The workload evaluates the ability to handle variable-length paths and multi-model filtering during traversal.

\subsection{Workload Summary}
Following the design principles above, the MAP-Bench workload suite comprises 19 read-only analytical queries. Derived from real-world academic search and recommendation scenarios~\cite{googlescholar,dblp,DBLP:conf/wsdm/0001A26,DBLP:journals/corr/abs-2512-06879,DBLP:journals/kais/ZhangPYZSRCW23}, these queries systematically vary operator combinations and data access paths across different data models to capture diverse system behaviors. The query descriptions are summarized in Table~\ref{tab:workload}, with technical dimensions detailed as follows:

\textbf{H-Pattern.} This category focuses on selective interactive lookups spanning multiple data models. H1 combines JSON containment predicates with graph traversal, while H4 retrieves neighbors via graph navigation and extracts nested authorship arrays. H2 targets relational-graph joins, linking nodes to structured metadata. H3 retrieves candidate papers through graph lookup and ranks them using edge attributes and vector distances. Finally, H5 retrieves nodes along the shortest citation path, extracts nested JSON attributes, and computes vector distances.

\textbf{A-Pattern.} The A-Pattern queries evaluate grouping, ranking, and aggregation over diverse multi-model access paths.  A1 and A4 first identify anchor entities by unnesting document arrays and aggregating (e.g., top institutions or hot topics), then expand through graph pattern matching to aggregate their related context. A2 targets aggregation over properties embedded in graph edges (e.g., co-authorship timelines). A3 and A6 evaluate aggregation performance over subsets filtered by upstream Graph (A3) or Vector (A6) operations. Finally, A5 targets multi-predicate filtering, intersecting results from a JSON containment query and a relational range scan before aggregation.

\textbf{V-Pattern.} The V-Pattern queries explore different placements of vector search within cross-model pipelines. V1 and V4 follow a \emph{Structure-First} (\emph{pre-filtering}) strategy, pruning candidates through Graph reachability or Document predicates before vector scoring. In contrast, V2 and V3 follow \emph{Semantic-First} (\emph{post-filtering}) strategies: V3 performs a broader vector search before applying relational and document filters, while V2 adds an exclusion pattern, requiring vector search while explicitly excluding graph neighbors.

\textbf{G-Pattern.} The G-Pattern queries explore different topological traversal strategies. G1 and G3 both involve multi-hop expansion, but differ in depth and purpose: G1 performs deeper traversal (2--4 hops) with structural pruning and vector-based scoring, while G3 focuses on shallower (0--2 hop) citation reachability with downstream metadata retrieval. G2 emphasizes path-level analytics by computing an influence score  along the shortest citation path. G4 targets fixed-length pattern matching by identifying papers that bridge two topic sets.

\begin{table}[t]
\centering
\small
\caption{Summary of MAP-Bench Workload Queries. Models are listed in the order of first appearance across query stages.}
\label{tab:workload}
\begin{tabular}{cp{5.6cm}c}
\toprule
\textbf{Query} & \textbf{Description} & \textbf{Models} \\
\midrule
H1 & Find authors by name variants and list publications & D-G-R \\
H2 & Find co-authors of a given author with their affiliations & G-R \\
H3 & Find papers belonging to a given topic, ranked by structural and semantic relevance & R-G-V \\
H4 & Find references of a paper with authors & G-D \\
H5 & Find intermediate papers on the citation path between two works, ranked by topic relevance & G-D-V \\
\midrule
A1 & Analyze top publishing institutions and their primary research fields & R-D-G \\
A2 & Analyze collaboration frequency for a given author, grouped by year & G-D-R \\
A3 & Analyze topic distribution for researchers at a given institution & R-G \\
A4 & Analyze which institutions lead in the most active research topic & R-D-G \\
A5 & Analyze prolific authors in a given topic whose papers contain specific keywords & R-D \\
A6 & Analyze top papers across related areas & V-G-R \\
\midrule
V1 & Recommend papers similar to a highly-cited seed paper within its citation network & R-D-G-V \\
V2 & Recommend novel papers beyond a paper's existing references & V-G \\
V3 & Recommend semantically similar papers, filtered by keywords and time range & V-R-D \\
V4 & Recommend papers similar to a given paper in a specified topic, with bibliographic details & R-D-V \\
\midrule
G1 & Explore potential collaborators in a specific research field & R-G-D-V \\
G2 & Explore the influence of papers along the shortest citation path between two works & G-R-D \\
G3 & Explore the citation neighborhood of a paper with author details & G-R-D \\
G4 & Explore interdisciplinary papers bridging two research fields & R-G-D \\
\bottomrule
\end{tabular}
\end{table}

\subsection{Evaluation Framework}
\label{sec:framework}

MAP-Bench considers three complementary aspects of cross-model pipelines: performance, resource, and query-writing complexity.

\textbf{Performance \& Resources.} Performance is measured as the median end-to-end query execution latency across multiple runs under both cold and hot conditions. Cold runs restart the DBMS and clear system cache before each query to ensure consistent initial states, whereas hot runs execute all queries consecutively for multiple rounds. To quantify resource usage, we record total CPU time (user + system) and peak resident memory usage ($\mathit{RSS_{peak}}$) during execution. Average CPU usage is approximated as $\mathit{time_{CPU}} / \mathit{latency}$, which reflects the average parallelism exploited during execution. We report this value as the average number of CPU cores utilized. Peak memory is obtained via periodic RSS sampling, and for multi-process systems, CPU time and memory usage are summed across all related processes.

\textbf{Complexity.} We also consider the complexity faced when writing cross-model queries. We approximate this developer-level complexity using Halstead metrics~\cite{halstead1977elements}, which quantify program complexity based on the counts of distinct and total operators and operands, yielding composite indicators such as \textit{volume} (information content) and \textit{effort} (estimated cognitive cost). These metrics have been adopted in prior studies to evaluate the conciseness and development effort of query languages in multi-model settings~\cite{DBLP:conf/er/ShiWL24,DBLP:conf/icde/Galvizo024}, and provide an additional perspective on the engineering cost of writing complex cross-model queries, complementing performance and resource evaluations.

\section{Evaluation}
\subsection{Systems Under Test}

We selected several representative systems that support the R/D/G/V models through different architectural approaches.

\textbf{AgensGraph}~\cite{agensgraph} is a PostgreSQL-based graph-relational database. Graph queries are expressed using Cypher syntax embedded in SQL. In our evaluation, document data are stored using the JSONB type and vector similarity search is enabled via pgvector extension.

\textbf{ArangoDB}~\cite{arangodb} is a native multi-model database supporting document and graph data within a unified engine. Queries are written in its native multi-model query language AQL, which provides integrated operators for document retrieval, graph traversal, and vector search. Relational data are represented using document collections.

\textbf{DuckDB}~\cite{DBLP:conf/sigmod/RaasveldtM19} is an in-process analytical database that, in our evaluation, uses its built-in JSON type together with the DuckPGQ extension for SQL/PGQ graph pattern matching and the VSS extension for vector similarity search.

\textbf{GredoDB}~\cite{GredoDB} is a multi-model research prototype built on openGauss. Its query language extends SQL with SQL/JSON-style path expressions for nested documents and graph patterns in the \texttt{FROM} clause, whose bound variables can be referenced in subsequent SQL clauses, enabling cross-model queries within a single statement. The execution engine provides native graph operators, while vector search is supported through pgvector-based integration. GredoDB is under active development and remains an early-stage prototype.

\textbf{Polyglot Baseline.} We construct a polyglot system from: PostgreSQL~\cite{postgresql} (R), MongoDB~\cite{mongodb} (D), Neo4j~\cite{neo4j} (G), and Milvus~\cite{milvus} (V), coordinated by a client-side middleware. The middleware applies two data-exchange strategies based on intermediate result size: \textit{(i)~Application-layer semi-join}—when intermediates are small, identifiers are embedded in \texttt{IN} predicates of downstream queries, with final joins completed in the middleware; \textit{(ii)~Cross-system materialization and pushdown}—when intermediates are large or require operators unsupported by the source engine, they are materialized in the target engine to leverage native joins or aggregations. For multi-column intermediates, only join keys are forwarded. Both strategies aim to minimize cross-engine data transfer, providing a representative polyglot baseline.
\subsection{Settings}

All experiments were conducted on a server with an Intel Xeon E7-8880 v4 CPU (176 cores), 1~TB memory, and 2~TB SSD storage, running CentOS 7. Each system was restricted to at most 16 CPU cores and 64~GB memory using system-level controls. For the Polyglot baseline, the 64~GB memory budget was split according to their memory requirements: PostgreSQL (10~GB), MongoDB (10~GB), Neo4j (22~GB), and Milvus (14~GB), with the remainder assigned to the middleware; Neo4j uses more for JVM heap, and Milvus for in-memory HNSW. Indexes were created on the same attributes whenever supported, using each system's native indexing mechanisms (e.g., varying support for indexing nested document fields and vector index types) to reflect practical multi-model capabilities. For vector search, index parameters were tuned to approximately $\text{recall@100} \approx 0.95$ to keep top-k results largely stable.

Hot and cold executions follow Section~\ref{sec:framework}; each hot-cache query was executed five times. To ensure result correctness and reproducibility, query parameters were kept identical across systems, guaranteeing consistent outputs for deterministic operations (R/D/G); approximate vector search (V) yields a largely stable top-k under our recall setting. The parameter settings yield non-trivial work across multiple models and bring out the choke points described in Section~\ref{sec:workload}. To facilitate reproducibility, system versions, indexing configurations, parameters, and scripts are available in our artifacts.

\subsection{Quality Evaluation of Generated Data}
\begin{figure*}[t]
    \centering
    \includegraphics[width=\textwidth]{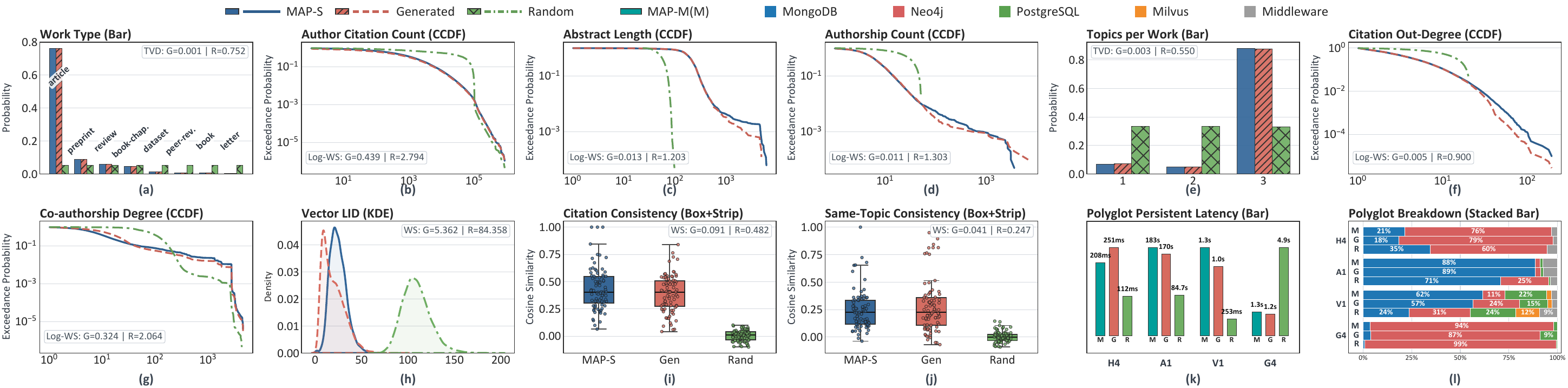}
    \caption{\textbf{Quality Evaluation of Generated Data.}
    (a)-(b) evaluate relational categorical and numerical distributions ; (c)-(e) evaluate document characteristics, including text length and nested-array cardinalities; (f)-(g) evaluate graph degree distributions; (h) evaluates vector-model LID; (i)-(j) evaluate cross-model semantic consistency; and (k)-(l) compare polyglot latency and per-model breakdown across datasets.}
    \label{fig:quality_overview} 
\end{figure*}

To validate the generation method, we evaluate both statistical distributions and performance fidelity. We compare real-world datasets (MAP-S, the generation seed, for distributional statistics, and MAP-M for query performance) against their counterparts generated by our method and by a random baseline constructed using the same procedure but without real-data statistics.  This design evaluates overall generation fidelity and the role of real-data-driven statistics in preserving these properties. In our experiments, with a configured parallelism of 72, the generator uses approximately 30 GiB of memory and generates about 12 papers per second. At the single-model level, we examine representative characteristics of each model (shown in Fig.~\ref{fig:quality_overview}); for the vector model, we additionally use local intrinsic dimensionality (LID) to characterize local structure~\cite{aumuller_role_2019}. At the cross-model level, we evaluate  semantic consistency across models by comparing vector similarities of sampled citation-linked and same-topic paper pairs. For performance fidelity, we measure polyglot end-to-end latency and per-model execution ratios. To quantify distributional differences, we use total variation distance (TVD)~\cite{gibbs2002choosingboundingprobabilitymetrics} for discrete categorical distributions, the first Wasserstein distance (WS)~\cite{gibbs2002choosingboundingprobabilitymetrics} for continuous distributions, and log-transformed Wasserstein distance (Log-WS) for heavy-tailed data after a $\log(1+x)$ transformation.

Experimental results are summarized in Fig.~\ref{fig:quality_overview}. Overall, Generated consistently outperforms Random and closely matches the Baseline. For single-model statistics (Fig.~\ref{fig:quality_overview}(a)-(h)), distributional distances (e.g., TVD and Wasserstein) are an order of magnitude lower than Random, indicating preserved structural and statistical properties. At the cross-model level, the WS values for citation and topic consistency (Fig.~\ref{fig:quality_overview}(i)-(j)) confirm that semantic relationships are also retained. Beyond statistical fidelity, we also evaluate performance consistency (Fig.~\ref{fig:quality_overview}(k)-(l)). End-to-end query latencies and per-model execution ratios on the generated data show comparable trends to the real-world baseline. In contrast, the random dataset, which lacks semantic alignment, exhibits noticeable performance deviations. This indicates that without semantic consistency, runtime execution patterns may vary significantly,  potentially leading to inaccurate performance conclusions. Overall, these results suggest that the generated data provides a reasonable approximation of real-world data for multi-model database benchmarking.
\subsection{Query Performance Analysis}
\label{sec:perf}

\begin{figure*}[t]
    \centering
    \includegraphics[width=0.95\textwidth]{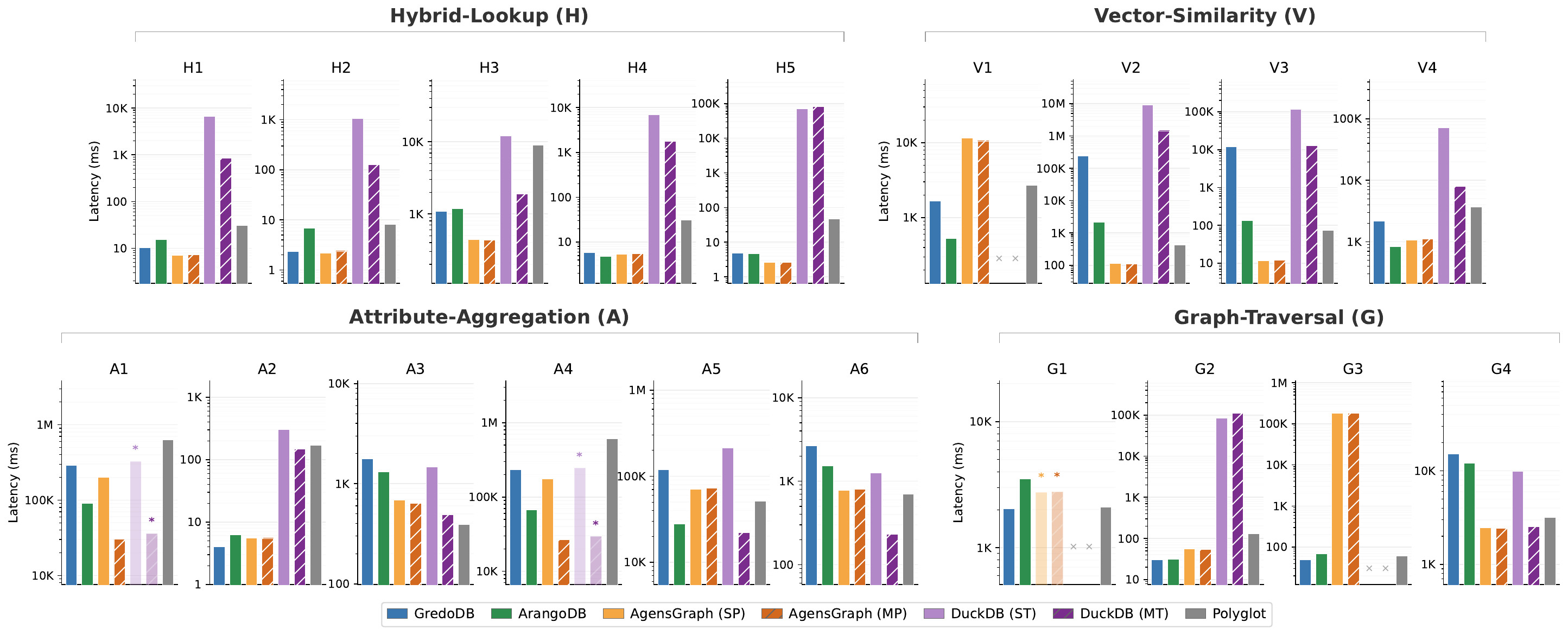}
    \caption{Query latency comparison under hot-cache execution.}
    \label{fig:perf_havg} 
\end{figure*}

\begin{figure}[h]
    \centering
    \includegraphics[width=0.8\linewidth]{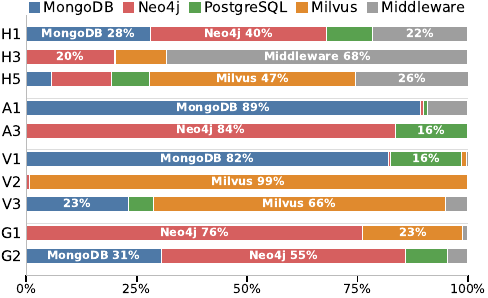}
    \caption{Latency breakdown in Polyglot.}
    \label{fig:poly_breakdown} 
\end{figure}

Figure~\ref{fig:perf_havg} shows hot-cache query latencies across H-A-V-G patterns on MAP-L, while Figure~\ref{fig:poly_breakdown} breaks down representative Polyglot executions into engine and middleware time. For AgensGraph, we report single-process (SP) and multi-process (MP) executions; for DuckDB, single-threaded (ST) and multi-threaded (MT) executions. Entries marked ``$\times$'' denote executions that did not complete in practice, while bars marked with ``*'' denote results obtained under relaxed settings and shown for reference only to preserve fairness in the main comparison. Unlike prior flat or domain-oriented benchmarks where execution characteristics are implicit and performance trends are scattered, MAP-Bench organizes queries by execution patterns and executes them as multi-stage pipelines, enabling the following direct observations.

\textbf{H-Pattern: Join strategy drives performance differences.}
Hybrid-Lookup queries start from highly selective predicates and perform small cross-model fetches, making them sensitive to index usage and join selection. AgensGraph performs best, benefiting from PostgreSQL's mature optimizer and diverse join strategies. GredoDB remains competitive through native graph navigation, while ArangoDB is slightly slower on index-driven lookups but faster on nested JSON extraction (H4); its plans are dominated by index nested-loop joins, which are effective for simple lookups but less flexible than PostgreSQL's join strategies. In contrast, DuckDB is consistently much slower; its documentation notes that ``ART indexes do not affect join performance,'' and its use of hash joins requires scanning full tables to build hash tables, even for selective queries. Additional overhead appears in shortest-path queries (H5), where the DuckPGQ constructs a full-graph CSR per query. For Polyglot, these lightweight queries are especially sensitive to middleware serialization and transfer overheads (Figure~\ref{fig:poly_breakdown}).

\textbf{A-Pattern: Parallelism dominates heavy aggregation.}
Aggregation queries are dominated by grouping and ranking over large inputs, making parallelism the key factor. AgensGraph-MP delivers the best performance on the heaviest queries (A1, A4), achieving 6–9$\times$ speedups over its single-process counterpart. DuckDB-MT likewise benefits from parallel execution by rewriting cross-model operations into parallelizable joins and scans, leading on A3, A5, and A6, but OOMing on A1/A4; with more relaxed $\sim$70GB peak RSS, it achieves latency close to AgensGraph-MP on A1/A4, showing the value of resource evaluation. ArangoDB performs well on complex document predicates (A5) but falls behind on large-scale scans due to the lack of parallel aggregation. GredoDB performs well on lightweight queries (A2) but scales poorly overall due to the absence of parallelism and feature limitations, as chained graph patterns are matched separately and then joined. An anomaly appears on A5: AgensGraph-MP shows no speedup, as the plan exposes no parallelism under complex JSON predicates, while DuckDB-MT remains efficient. In contrast, AgensGraph gains little on A3 and A6, as its graph traversal and vector search stages rely on index-based scans that do not benefit from parallel execution. The Polyglot baseline performs worst on A1 and A4, where most time is spent in MongoDB joining and aggregating intermediate results from PostgreSQL. On A3, however, Polyglot performs best because only two models are involved and intermediates remain small.

\textbf{V-Pattern: Pipeline position determines the bottleneck.}
For vector-similarity queries, the position of vector search determines the bottleneck. In \emph{pre-filtering} scenarios (V1, V4), execution is dominated by upstream operations. ArangoDB performs best on V1 and V4 due to efficient document filtering and variable-length expansion (VLE) traversal, while AgensGraph is slowed by its VLE operator. In \emph{post-filtering} scenarios (V2, V3), the bottleneck shifts to vector index quality: AgensGraph achieves the best V3 latency with HNSW, outperforming ArangoDB's IVF-FLAT. GredoDB is moderate in pre-filtering but weak in post-filtering, as vector index scans incur high cost in their plans. DuckDB cannot complete V1 due to limited VLE support; V4 is slowed by JSON extraction from string-based storage; and in V2–V3, its VSS extension requires a constant query vector and cannot accept subquery inputs, which limits cross-model pipelines and degrades to brute-force scanning. The Polyglot baseline performs well when vector search dominates but is penalized when upstream operations become the bottleneck, a shift dependent on vector placement that prior benchmarks without cross-model vector integration cannot capture.

\textbf{G-Pattern: Dynamic vs. static path execution drives performance.}
Graph-traversal queries exhibit the largest performance variation. On \emph{variable-length} traversal (G1, G3), AgensGraph shows divergent behavior: it is orders of magnitude slower than GredoDB and ArangoDB on G3 due to inefficient variable-length expansion (VLE), while on G1 its native operator does not complete in practice. A fixed-length UNION rewrite runs in a few seconds, suggesting that the bottleneck lies in native dynamic path execution. DuckDB also struggles with variable-length queries (G1, G3), as DuckPGQ translates them into full-graph CSR construction and large iterative self-join chains, leading to plans that do not complete in practice; shortest-path queries (G2) are supported but slow for the same reason. On \emph{fixed-length} pattern matching (G4), AgensGraph and DuckDB-MT are both competitive, indicating that the bottleneck lies in dynamic path execution rather than graph operations in general. GredoDB achieves the best performance on dynamic traversal queries (G1–G3), but is weaker on fixed-length matching (G4) due to prototype limitations. ArangoDB performs well on shortest-path queries but is slightly slower on fixed-length patterns. The Polyglot baseline performs consistently well on traversal queries, but incurs additional overhead on G2 due to document lookups after traversal.

\textbf{No single system dominates across all patterns.} MAP-Bench shows that strengths are pattern-dependent: AgensGraph is strong on interactive lookups and some analytical queries; DuckDB-MT excels at parallel aggregation but struggles on traversal and interactive queries; ArangoDB performs best on multi-model pre-filtering vector and document-oriented workloads but lacks aggregation parallelism; GredoDB is strong on graph traversal but limited by its prototype implementation; and Polyglot is competitive when one engine dominates but incurs overhead on cross-model pipelines.

\begin{figure}[t]
    \centering
    \includegraphics[width=0.9\linewidth]{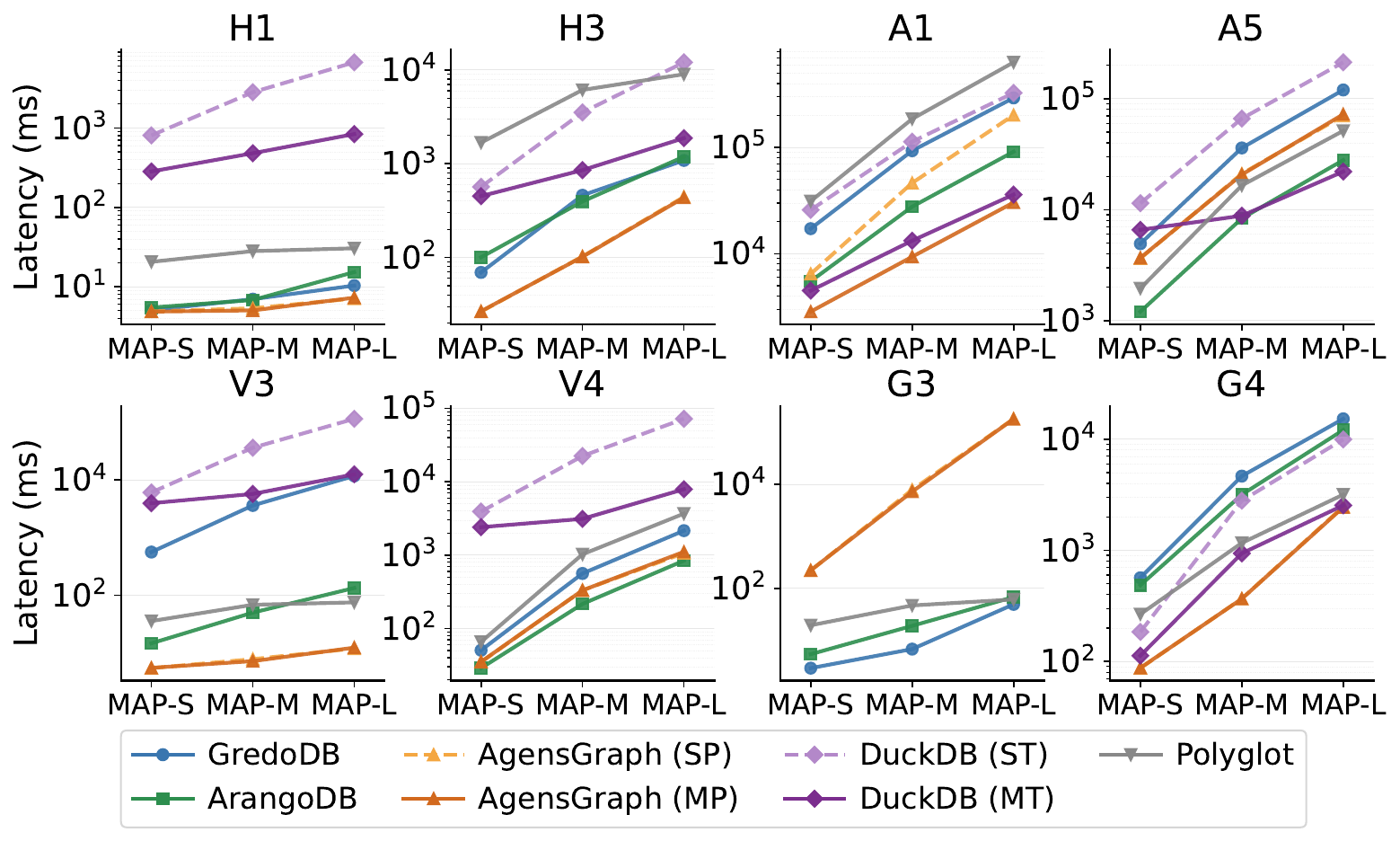}
    \caption{Scalability of representative H-A-V-G queries.}
    \label{fig:scalability} 
\end{figure}

Figure~\ref{fig:scalability} shows scalability under MAP-Bench from MAP-S to MAP-L for representative H-A-V-G queries (with parameters adjusted per scale to keep workloads non-trivial). Overall, varying dataset scale reinforces the bottlenecks identified in the main evaluation, while leaving the qualitative conclusions unchanged, although a small subset of queries shows ranking shifts with scale. In the H and V patterns, H1/H3 and V3/V4 largely preserve their relative ordering across MMDB systems, indicating stable dominant bottlenecks. In the aggregation pattern, A1 continues to favor parallel engines such as AgensGraph-MP and DuckDB-MT, whereas A5 shows a different trend: DuckDB-MT is less competitive at small scale, where JSON parsing overhead is more visible, but improves at larger scales as parallel scan and aggregation amortize this cost. Polyglot shows a similar trend on H3, G3, and V3: end-to-end latency is more sensitive to the fixed cost of middleware round trips at small scale, while this cost accounts for a smaller share as engine-side execution grows. In the graph pattern, G3 and G4 separate dynamic traversal from fixed-length matching, with GredoDB stronger on dynamic traversal and AgensGraph more competitive on fixed-length graph patterns.

\begin{table}[t]
\caption{Geometric Mean Cold-To-Hot Speedup Ratios.}
\label{tab:cold_hot}
\centering
\resizebox{0.45\textwidth}{!}{  
\begin{tabular}{|c|c|c|c|c|c|}
\hline
\textbf{System}        & \textbf{H}    & \textbf{A}    & \textbf{V}    & \textbf{G}    & \textbf{ALL}   \\ \hline
\textbf{AgensGraph(MP)} & 4.8          & 1.51         & 1.91         & 1.47         & 2.14          \\ \hline
\textbf{ArangoDB}      & 4.89         & 2.21         & 2.59         & 4.21         & 3.22          \\ \hline
\textbf{DuckDB(MT)}    & 1.57         & 1.35         & 1.35         & 1.24         & 1.4           \\ \hline
\textbf{GredoDB}       & 1034.34      & 10.99        & 1.84         & 65.32        & 36.31         \\ \hline
\textbf{Polyglot}      & 78.91        & 3.79         & 10.58        & 13.1         & 13.58         \\ \hline
\end{tabular}
}
\end{table}

Table~\ref{tab:cold_hot} reports the geometric mean of cold-to-hot latency ratios per pattern. DuckDB-MT exhibits the most stable performance, owing to its in-process, analytical architecture, where query latency is largely independent of cache state. Both AgensGraph-MP and ArangoDB show moderate cold-to-hot ratios, with the H-pattern being the most affected due to its reliance on random index lookups that benefit from hot caches. GredoDB shows the highest cold-to-hot ratios on the H and G patterns, as its prototype architecture requires loading the full graph topology from disk on the first graph query. The Polyglot baseline shows elevated cold-to-hot gaps, as its independent engines must warm up separately; the effect is most pronounced on H queries due to cross-engine round trips. Across all systems, the A-pattern is the least affected by cache state, as its scan-dominated execution relies on sequential I/O rather than cached index lookups.

\subsection{Resource Analysis}

\begin{table}[h]
\centering
\caption{Resource usage of representative queries. Each cell reports (avg CPU cores, peak RSS in GB).}
\label{tab:resource_usage}
\setlength{\tabcolsep}{2pt}
\resizebox{0.45\textwidth}{!}{
\begin{tabular}{lccccc}
\toprule
\textbf{Q} & \textbf{AgensGraph(MP)} & \textbf{ArangoDB} & \textbf{DuckDB(MT)} & \textbf{GredoDB} & \textbf{Polyglot} \\
\midrule
H1 & (1.3, 0.02) & (1.1, 1.02) & (13.9, 1.10) & (1.1, 9.89) & (12.4, 27.4) \\
H3 & (1.0, 0.69) & (1.0, 1.88) & (13.4, 13.5) & (1.1, 5.47) & (5.1, 29.1) \\
H5 & (1.0, 0.01) & (1.1, 1.09) & (2.1, 46.8) & (1.1, 4.04) & (7.1, 29.0) \\
\midrule
A1 & (5.1, 32.7) & (1.0, 27.9) & OOM & (1.2, 62.31) & (1.5, 35.1) \\
A4 & (5.2, 32.6) & (1.0, 28.7) & OOM & (1.2, 58.28) & (1.4, 35.0) \\
A5 & (1.0, 15.2) & (1.0, 23.0) & (13.7, 28.3) & (1.2, 5.40) & (2.8, 33.1) \\
\midrule
V3 & (1.1, 0.03) & (1.1, 1.12) & (13.9, 40.2) & (1.1, 1.73) & (2.1, 27.2) \\
V4 & (1.0, 0.63) & (1.0, 2.04) & (13.7, 38.8) & (1.4, 1.71) & (5.5, 31.6) \\
\midrule
G2 & (1.0, 0.08) & (1.1, 0.96) & (2.1, 46.8) & (1.1, 3.98) & (12.0, 27.4) \\
G3 & (1.0, 2.89) & (1.1, 1.00) & -- & (1.1, 3.97) & (12.7, 27.3) \\
\bottomrule
\end{tabular}
}
\end{table}

Table~\ref{tab:resource_usage} reports resource usage measured by MAP-Bench on representative queries from each H-A-V-G pattern.

\textbf{CPU.} The CPU results reinforce the observations in Section~\ref{sec:perf}. DuckDB exploits the most parallelism, using 13--14 cores on most queries; only shortest-path queries (H5, G2) drop to about 2 cores due to DuckPGQ limitations. AgensGraph applies parallelism more selectively, reaching about 5 cores on the heavy aggregations A1/A4 but remaining near single-core execution elsewhere. ArangoDB and GredoDB stay close to 1 core on nearly all queries, consistent with their limited intra-query parallelism. Polyglot also shows high CPU usage (5--13 cores), but this reflects the summed cost of multiple engines and middleware rather than a single parallel execution engine. On A1 and A4, however, utilization drops to about 1.5 cores because MongoDB aggregation becomes the bottleneck and the rest of the pipeline must wait for its output.

\textbf{Memory.}
From the resource data, AgensGraph and ArangoDB generally show the lowest RSS on most representative queries, with noticeable growth mainly on the heaviest aggregations. DuckDB shows the highest memory demand: even several non-aggregation queries require tens of GB, and A1/A4 further exceed the default 64~GB budget, consistent with its in-memory parallel execution model. GredoDB uses moderate memory on most queries, but shows elevated RSS on A1/A4, likely due to prototype-stage memory overheads. Polyglot consistently shows high RSS across all queries, largely because Neo4j's JVM heap remains resident after startup.

\subsection{Query Complexity Analysis}

\begin{table}[h]
\centering                                                                                                                            
\caption{Comparison of cross-model query writing complexity using Halstead metrics.}
\label{tab:full_metrics}
\setlength{\tabcolsep}{2pt}
\resizebox{0.45\textwidth}{!}{
\begin{tabular}{lccccccccc}
\toprule
\multirow{2}{*}{\textbf{System}} 
& \multicolumn{3}{c}{\textbf{Volume}} 
& \multicolumn{3}{c}{\textbf{Difficulty}} 
& \multicolumn{3}{c}{\textbf{Effort}} \\
\cmidrule(lr){2-4} \cmidrule(lr){5-7} \cmidrule(lr){8-10}
& Avg & Max & Std & Avg & Max & Std & Avg & Max & Std \\
\midrule
GredoDB    & 978   & 1,846 & \textbf{378} & 33.2 & 51.7 & \textbf{8.1} & 34,888  & 95,479  & \textbf{21,323} \\
ArangoDB   & \textbf{960} & \textbf{1,729} & 425 & \textbf{30.8} & \textbf{49.6} & 8.4 & \textbf{32,697} & \textbf{85,771} & 22,664 \\
AgensGraph & 1,122 & 3,085 & 596 & 37.8 & 72.2 & 12.1 & 48,741  & 222,682 & 48,137 \\
DuckDB     & 1,101 & 1,820 & 435 & 37.2 & 56.4 & 10.5 & 44,734  & 102,744 & 26,973 \\
Polyglot  & 2,614 & 4,480 & 1,082 & 47.0 & 77.8 & 15.0 & 137,145 & 348,655 & 93,898 \\
\bottomrule
\end{tabular}
}
\end{table}

Following Section~\ref{sec:framework}, we compute Halstead metrics~\cite{halstead1977elements} over all queries as an approximate proxy for developer-facing cross-model query complexity. To improve comparability across heterogeneous query languages, we use a unified tokenizer that identifies operators and operands while filtering comments and irrelevant literals. Table~\ref{tab:full_metrics} summarizes the results.

\textbf{Unified multi-model languages.}
ArangoDB and GredoDB impose the lowest query-writing burden overall. ArangoDB has the lowest average effort, reflecting AQL’s uniform \texttt{FOR}--\texttt{FILTER}--\texttt{RETURN} structure across document, graph, and vector operations. GredoDB is close behind and shows the lowest variability, because graph patterns are embedded directly in the \texttt{FROM} clause and their variables can be referenced in standard SQL clauses without extra wrappers or projection layers. In both systems, developers can stay within a relatively uniform query model rather than repeatedly switching between different abstractions.

\textbf{Wrapper-based graph integration.}
AgensGraph and DuckDB require about $1.4\times$ more effort than ArangoDB on average. In both systems, graph logic must first be expressed in a separate construct and then re-integrated into SQL for subsequent processing. This adds extra query boundaries, intermediate structure, and context switches, especially in multi-stage queries that combine traversal with relational aggregation.

\textbf{Polyglot orchestration.}
The Polyglot baseline imposes the highest query-writing burden, requiring over $4\times$ the effort of Arango\-DB on average and showing far greater variability. Here, the burden comes not only from longer query expressions, but from coordinating multiple query languages together with application-side orchestration code. As a result, part of the cross-model integration effort is shifted from the database engine to the developer.

\subsection{Insights from Evaluation}
\label{sec:insights}

Our evaluation provides several insights into the design and optimization of multi-model database systems.

\textbf{Cross-model query optimization.}
Interactions among operations across different data models in multi-stage pipelines can shift performance bottlenecks and affect end-to-end query performance. Beyond optimizing individual data models, query optimization should also consider the entire cross-model pipeline, accounting for factors such as operator dependencies, intermediate result sizes, and data movement costs.

\textbf{Parallel execution.}
Parallelism significantly accelerates analytical queries, but insufficient parallel support for certain multi-model operators limits its benefits. Improving parallel execution capabilities across different data models while controlling memory overhead remains an important direction.

\textbf{Unified Query Languages.}
Unified multi-model languages generally require less estimated query-writing effort than Polyglot orchestration. Allowing heterogeneous operations to be expressed and composed within a unified query language can reduce explicit cross-model coordination and the cognitive burden of switching between different query languages.

\section{Conclusion}
\label{sec:conclusion}

We presented MAP-Bench, a benchmark for cross-model analytical workloads spanning relational, document, graph, and vector models. Built on a semantically consistent real-world dataset from OpenAlex, it organizes workloads by execution patterns (H-A-V-G) and evaluates systems across latency, resource consumption, and query complexity. By combining real-world semantic consistency with pattern-oriented cross-model pipelines, particularly the integration of vector semantic search missing from prior benchmarks, MAP-Bench enables systematic analysis of system behavior across query patterns. Our evaluation shows that no single system dominates across all patterns, revealing distinct performance bottlenecks and providing insights into cross-model query optimization, parallel execution, and query language design.

\section*{Artifact Availability}
The MAP-Bench datasets, data generator, workload implementations,
system adapters, and scripts for reproducing the experiments are
publicly available at
\url{https://github.com/user-kk/MAP-Bench}.




\bibliographystyle{ACM-Reference-Format}
\bibliography{references}


\end{document}